\documentclass{cimento}

\usepackage{graphicx}  
\usepackage{booktabs}
\usepackage{upgreek}
\usepackage{url}
\usepackage{physics}
\usepackage{xpatch}
\usepackage{xspace}
\usepackage{caption}
\usepackage[list=true,
labelformat=brace, position=top]{subcaption}
\usepackage[separate-uncertainty=true]{siunitx}
\usepackage{hepnames}
\usepackage{xpatch}
\usepackage{xspace}

\makeatletter
\xpatchcmd\@HepConStyle
 {\edef\@upcode{\updefault}}
 {\ifdefined\shapedefault\edef\@upcode{\shapedefault}\else\edef\@upcode{\updefault}\fi}
 {}{}
\makeatother

\newcommand{\threepi}{\ensuremath{\Ppiminus\Ppiminus\Ppiplus\Pnut}\xspace}
\newcommand{\tauthreepi}{\ensuremath{\Ptauon\to\threepi}\xspace}
\newcommand{\tautothreepi}{\tauthreepi}
\newcommand{\tautofourpi}{\ensuremath{\Ptauon\to\Ppiminus\Ppiminus\Ppiplus\Ppizero\Pnut}\xspace}
\newcommand{\eeqq}{\ensuremath{\APelectron\Pelectron\to\Pquark\APquark}\xspace}
\newcommand{\eett}{\ensuremath{\APelectron\Pelectron\to\APtauon\Ptauon}\xspace}

\newcommand{\aots}{\ensuremath{\mathrm{a}_1(1260)}\xspace}
\newcommand{\aosf}{\ensuremath{\mathrm{a}_1(1640)}\xspace}
\newcommand{\aoft}{\ensuremath{\mathrm{a}_1(1420)}\xspace}
\newcommand{\rhoss}{\ensuremath{\rho(770)}\xspace}
\newcommand{\fnotfh}{\ensuremath{\mathrm{f}_0(500)}\xspace}
\newcommand{\fnotne}{\ensuremath{\mathrm{f}_0(980)}\xspace}
\newcommand{\fnotfth}{\ensuremath{\mathrm{f}_0(1500)}\xspace}
\newcommand{\ft}{\ensuremath{\mathrm{f}_2(1270)}\xspace}
\newcommand{\rhoprime}{\ensuremath{\rho(1450)}\xspace}
\newcommand{\omegase}{\ensuremath{\omega(782)}\xspace}

\newcommand{\pipiS}{\ensuremath{[\pi\pi]_\mathrm{S}}\xspace}

\newcommand{\fnotpiP}{\ensuremath{1^+[\mathrm{f}_{0}(980)\pi]_\mathrm{P}}\xspace}

\newcommand{\sigmapiP}{\ensuremath{1^+[\sigma\pi]_\mathrm{P}}\xspace}
\newcommand{\rhopiS}{\ensuremath{1^+[\rho(770)\pi]_\mathrm{S}}\xspace}

\title{Partial-wave analysis of \tautothreepi{} at BELLE}
\author{A.~Rabusov\from{ins:TUM},
D.~Greenwald\from{ins:TUM},
\atque
S.~Paul\from{ins:TUM}
for the Belle collaboration
}
\instlist{\inst{ins:TUM} Technical University of Munich,
James-Franck str. 1, 85748 Garching, Germany}

\begin{document}

\maketitle
\begin{abstract}
    We present preliminary results of a partial-wave analysis of
    \tauthreepi in data from the Belle experiment at the KEKB
    $\APelectron\Pelectron$ collider. We demonstrate the presence of
    the \aoft and \aosf resonances in tauon decays and measure their
    masses and widths. We also present validation of our findings
    using a model-independent approach. Our results can improve
    modeling in simulation studies necessary for measuring the tauon
    electric and magnetic dipole moments and Michel parameters.
\end{abstract}

Studies of the correlated spins of tauons in \eett, such as measurement
of the tauon electric and magnetic dipole moments, look at events in
which tauons decay to \threepi\footnote{We imply the charge-conjugated
mode throughout this text.}~\cite{b2pb, krinner-edm}.  However, lack of
knowledge about the dynamics of \tautothreepi, which have never been
experimentally measured in detail, limits the sensitivity of such
measurements.

This decay proceeds mostly through the \aots resonance, a ground state
unflavoured axial-vector meson~\cite{pluto} whose mass and width are
poorly known~\cite{compass-a1,cleo-ii-tau3pi,pdg2022}. Also poorly
understood is what other resonances are present and in which admixture
they contribute. The COMPASS experiment observed a
narrow, unflavoured axial-vector resonance, \aoft, in pion-proton
scattering~\cite{compass-a1-1420}. If it exists, it should be present
in \tauthreepi. Whether it is a particle or an artifact of
$\PKstar\PK$ scattering can be clarified by studying it in tauon
decay~\cite{balalaika}

We perform a partial-wave analysis~(PWA) of \tautothreepi using
\SI{980}{fb^{-1}} of data collected by the Belle experiment at the
asymmetric $\APelectron\Pelectron$ collider KEKB.


\tautothreepi is parameterized by seven phase-space variables: the mass
of three pions, $m_{3\Ppi}$; the squared masses of the two
$\Ppiplus\Ppiminus$ pairs, $s_1$ and $s_2$; the three Euler angles of
the decay plane; and the helicity angle of the neutrino in the tauon
rest frame~\cite{kuhn}. One of the Euler angles is immeasurable because
we cannot detect the \Pnut. We average the decay rate over this angle.
We studied the effect of the averaging procedure and found it introduces
no biases~\cite{ichep2022}.

Our event selection criteria~\cite{ichep2022} are optimized, using
simulated data, to maximize efficiency and purity, but chosen not to
distort the phase-space variables. In each event, we define two
hemispheres defined by the plane perpendicular the thrust axis of all
charged particles and photons. We require three charged particles be
present in one hemisphere, the signal hemisphere, and one be present in
the other, the tag hemisphere.

Using simulated data, we train a boosted decision tree~(BDT) to remove
events from processes other than \eett. It looks at six event
variables: the thrust, the sum of charged-particle and photon momenta
in the center-of-momentum~(CM) frame, the mass and cosine of the polar
angle of the missing four-momentum in the CM frame, the total energy
measured in the Belle's electromagnetic calorimeter, and the sum of the
energies of charged particles in the CM frame.  The thrust is the most
discriminating of the variables.

We veto the presence of charged kaons in the signal hemisphere by
requiring the particles with the same charge be consistent with being
pions. We veto the presence of neutral kaons by requiring
$|\sqrt{s_i} - m_{\PKzero}| > \SI{12}{MeV}$, $i=1,2$.
Finally, we veto the presence of neutral pions by requiring the sum of
photon energies in the signal hemisphere to be below \SI{480}{MeV}.

We select \num{55e6} events, with a signal-decay efficiency of 32\%
and a purity of 82\%. This is the largest sample of \tautothreepi to
date.  However, it is contaminated with \num{10e6} background events,
mostly originating from \tautofourpi and \eeqq, where $\Pquark$ can be a
$\Pup$-, $\Pdown$-, $\Pstrange$-, or $\Pcharm$-quark. To account for it,
we train a neural network on simulated data to describe the
multidimensional distribution of the background as described
in~\cite{polu-nn,ichep2022}. We add the neural network's prediction of
the background intensity to our partial-wave analysis.

We split the data into bins of $m_{3\Ppi}$ and perform a partial-wave
decomposition independently in each bin using the isobar model and
tensor formalism described in~\cite{krinner}. It assumes that the
decay proceeds through a resonance $\mathrm{X}^-$ that decays to three
charged pions via a sequence of two-body decays,
$\mathrm{X}^-\to\xi^0\Ppiminus$ and $\xi^0\to\Ppiminus\Ppiplus$, where
$\xi^0$ is an isobar. We ignore the nature of the resonance
$\mathrm{X}^-$ in the partial-wave decomposition, but we require it
have $J^P=1^+$, $0^-$, or $1^-$ quantum numbers. In the \tautothreepi
process, a decay amplitude with $J^P=1^-$ violates $G$ parity.

For $\xi^0$ we include \rhoss, \rhoprime, \fnotfh, \fnotne, \fnotfth,
\ft, and \omegase. We model all but \fnotfh with the relativistic
Breit-Wigner function with masses and widths taken from~\cite{rootpwa}
and \fnotfh with the broad \pipiS component described
in~\cite{kachaev}. We allow angular momenta between $\xi^0$ and the
remaining pion in the $[0,3]$ range.

\begin{figure}[!t]
    \begin{center}
        \begin{subfigure}[b]{0.48\textwidth}
            \includegraphics[width=\textwidth]{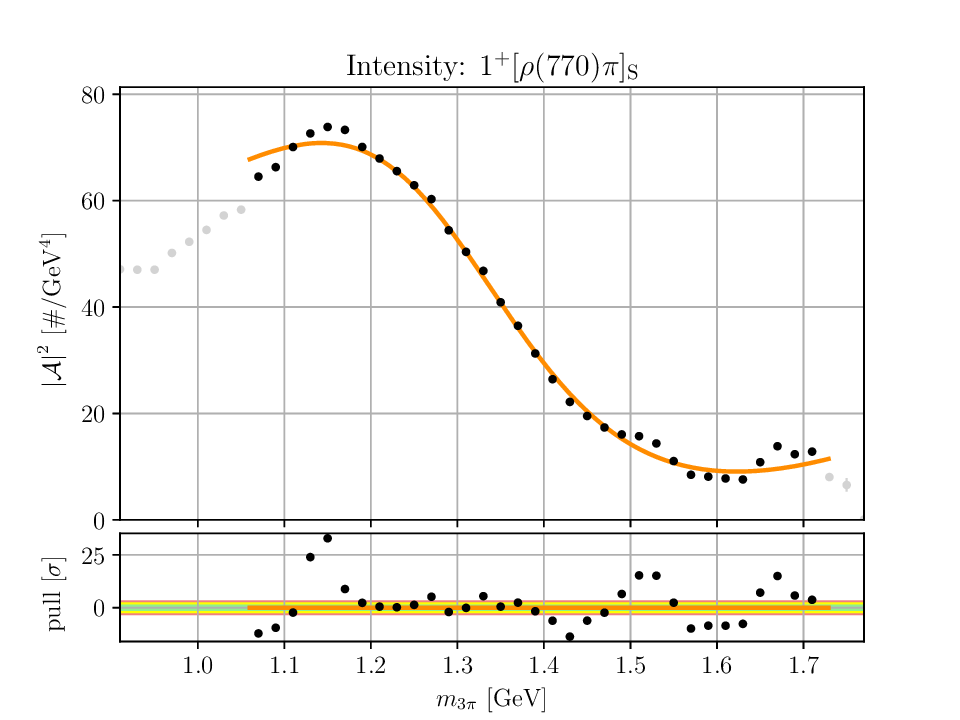}
            \caption{\rhopiS{}~wave.}
            \label{fig:a1_1260}
        \end{subfigure}
        \begin{subfigure}[b]{0.48\textwidth}
            \includegraphics[width=\textwidth]{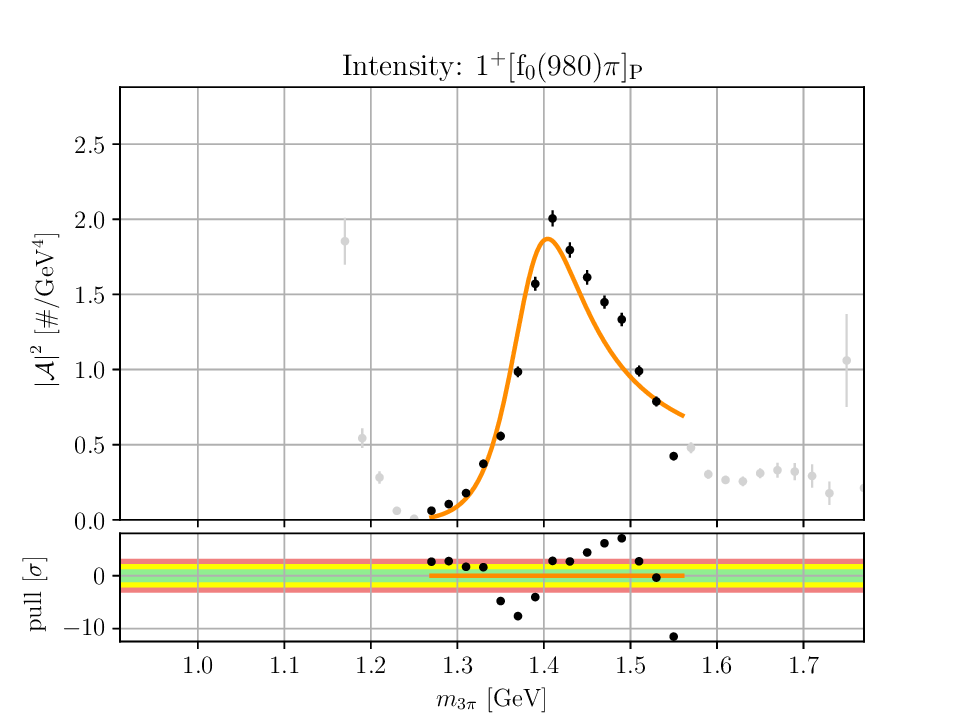}
            \caption{\fnotpiP{}~wave.}
            \label{fig:a1_1420}
        \end{subfigure}
        \caption{Intensities as functions of $m_{3\pi}$ for the
          partial-wave decompositions~(black and grey points), with
          statistical uncertainties shown, and the prediction of the
          resonance-model fit~(orange lines) to the black points.}
        \label{fig:mdf}
    \end{center}
\end{figure}

We observe the \rhopiS wave shown in Fig.~\ref{fig:a1_1260}, which has
a fit fraction around 100\%. The second most present partial wave is
the \sigmapiP wave with a fit fraction around 20\%. Fit fractions need
not sum to unity due to the interferences between partial waves. Our
results agree with those of CLEO~II for
$\Ptauon\to\Ppiminus\Ppizero\Ppizero\Pnut$ in~\cite{cleo-ii-tau3pi}.

We fit a model for the $\mathrm{X}^-$ to the results of the partial-wave
decomposition, accounting for the phase-space volume and efficiency in
each $m_{3\Ppi}$ bin. We use the Bowler parameterization~\cite{bowler}
for \aots and a relativistic Breit-Wigner function for \aosf and \aoft.
In the resonance-model fit shown in Fig.~\ref{fig:mdf}, we measure the
masses and widths of \aots and \aoft, which are shown in
Table~\ref{tab:res}. The mass and width of \aots agree with the COMPASS
results within their systematic uncertainties.  Fig.~\ref{fig:a1_1420}
demonstrates the fit projection of the \fnotpiP wave. The \aoft
resonance appears as a narrow peak near \SI{1400}{MeV}. Its mass and
width are smaller than those measured by COMPASS. This could be
explained by the presence of coherent background in COMPASS which
dominates the low $m_{3\pi}$ region.

\begin{table}[!t]
  \caption{Resonance-model fit results (statistical uncertainties only).}
  \label{tab:res}
  \begin{tabular}{lSS}
    \toprule
    
    & {mass [\si{MeV}]}
    & {width [\si{MeV}]} \\
    
    \aots
    & 1328.9 +- 0.1
    & 388.4 +- 0.1
    \\
    
    \aoft
    & 1387.8 +- 0.3
    & 109.2 +-0.6
    \\
    
    \bottomrule
  \end{tabular}
\end{table}

For the fit to converge, we fix the mass and width of \aosf to
the nominal values from the PDG~\cite{pdg2022} because of the
phase-space suppression in the region of the \aosf. We justify the
inclusion of \aosf by the intensity excess at the high-mass tail of
the \aosf resonance in the \rhopiS~intensity plot in
Fig.~\ref{fig:a1_1260}.

We verify the presence of \aoft using the quasi-model-independent PWA
technique developed in~\cite{zm}. At the partial-wave decomposition
stage, we parameterize the \pipiS wave with a complex-valued step-like
function and let the fit find the \pipiS wave amplitude as a function
of $m_{2\Ppi}$. To reduce the bias caused by the \rhopiS model, we
similarly parameterize the \rhopiS wave. We observe the presence of
the \fnotne isobar in the \pipiS wave in the narrow region of
$m_{3\Ppi}$ and extract its amplitude in the isobar fit. The
amplitude's intensity and phase have similar features to those from
the conventional PWA.

We conducted a preliminary study of the systematic uncertainties on
20\% of the data. We varied the neural network coefficients to study
the uncertainty on the background parameterization and find that its
tantamount to the statistical uncertainty for $m_{3\pi} >
\SI{1}{GeV}$.

In conclusion, we observe the \aoft resonance in both conventional and
quasi-model-independent PWA. After finalizing the systematic
uncertainty studies, we will provide an updated model for the
\tautothreepi with 15 partial waves, which can be used by the TAUOLA
software library for simulating tauon decay and in future measurements
of the tau electrical and magnetic dipole moments at Belle~II.

\acknowledgments

We acknowledge Florian Kaspar for providing his code to train a neural
network with the simulated background sample, Dmitrii Ryabchikov for
providing his code to resolve the ambiguities in the
quasi-model-independent PWA, and Stefan Wallner for cross-checking the
de-weighting scheme used for acceptance correction.

\end{document}